\documentstyle[12pt,epsf,epsfig,a4]{article}
\begin{document}
\begin {titlepage}
\begin{flushleft}
FSUJ TPI QO-10/97
\end{flushleft}
\begin{flushright}
November, 1997
\end{flushright}
\vspace{20mm}
\begin{center}
{\Large {\bf Three-dimensional quantization of the electromagnetic field in 
dispersive and absorbing inhomogeneous dielectrics} \\[3ex]
\large Ho Trung Dung$^*$, Ludwig Kn\"oll, and Dirk-Gunnar Welsch}\\[2.ex]
Friedrich-Schiller-Universit\"at Jena 
Theoretisch-Physikalisches Institut \\[1ex]
Max-Wien Platz 1, D-07743 Jena, Germany
\vspace{25mm}
\end{center}
\begin{center}
\bf{Abstract}
\end{center}
A quantization scheme for the phenomenological Maxwell theory of the 
full electromagnetic field in an inhomogeneous three-dimensional, 
dispersive and absorbing dielectric medium is developed. The classical 
Maxwell equations with spatially varying and Kramers--Kronig consistent 
permittivity are regarded as operator-valued field equations, introducing 
additional current- and charge-density operator fields in order to take 
into account the noise associated with the dissipation in the medium.
It is shown that the equal-time commutation relations between the fundamental 
electromagnetic fields $\hat{\bf E}$ and $\hat{\bf B}$ and 
the potentials $\hat{\bf A}$ and $\hat\varphi$ in the Coulomb gauge
can be expressed in terms of the Green tensor of the classical problem. From 
the Green 
tensors for bulk material and an inhomogeneous medium consisting of two 
bulk dielectrics with a common planar interface it is explicitly proven 
that the well-known equal-time commutation relations of QED are preserved.
\end{titlepage}
\renewcommand {\thepage} {\arabic{page}}
\setcounter {page} {2}

\section{Introduction}
\label{sec1}

It is well known that the quantum statistical properties of  
electromagnetic fields including their interactions with atomic 
systems can be strongly influenced by the presence of dielectric bodies.
Typical examples are the Casimir effect \cite{1,2}, the
modification of the spontaneous emission rate of excited
atoms \cite{2a,3,4} in the presence of dielectric media,
and the degradation or improvement of nonclassical properties of 
light propagating through optical devices, such as cavities, beam 
splitters, wave guides etc., which typically can be regarded as 
dielectric bodies\cite{5,6,7}. Therefore it has been of considerable 
interest to formulate QED on a dielectric-matter background. Various 
quantization schemes have been proposed for dispersionless
\cite{8,9,10,11,12,13,14,15}, dispersive~\cite{16,17,18,19,20,21,22}, 
and nonlinear~\cite{17,18,19,20,21,23,24,25} 
dielectrics. However, most of these quantization schemes run into 
difficulties when 
an absorbing medium is attempted to be included in the concept, 
which is crucial for studying propagation effects and keeping 
the theory consistent with the causality principle.

The problem has been considered by a number of  
authors \cite{30,31,32,33,34,34a,34b,35,37,37a,38,38a}. In \cite{32} 
a fully canonical quantization scheme for the macroscopic electromagnetic 
field in a linear harmonic-oscillator bulk material is developed which is
based on the Hopfield model of a dielectric \cite{36}. The electromagnetic 
field is coupled to a harmonic-oscillator polarization field that 
interacts with a continuum of harmonic-oscillator reservoir fields. 
The resulting Hamiltonian, which is a bilinear form of bosonic fields, 
is diagonalized in two steps -- first the polarization-reservoir part and 
after that the total Hamiltonian. 
The scheme is much more involved when it allows the electromagnetic
field to be in 
an inhomogeneous medium, as it is the case in practice,
and much effort must be made to perform the diagonalization even
for simple dielectric-body configurations \cite{38}.

Another approach to the problem of including losses in the quantization 
scheme is the method of Green function expansion \cite{39,40}, which 
can be regarded as a natural extension of the familiar method
of mode expansion (which only applies to strictly nonabsorbing media)
to arbitrary Kramers--Kronig consistent media. The approach, which resembles,
in a sense, the method of (operator) Langevin forces \cite{33,34,40a,40b},
directly starts with the Maxwell equations for the macroscopic 
electromagnetic field, including the dielectric displacement vector and  
a (phenomenologically) given permittivity. 
The quantization of the radiation field is based on the classical 
Green function representation of the vector potential, identifying the 
external sources therein with the noise sources that are necessarily 
associated with the losses in the medium and replacing the $c$-number 
sources with operator-valued ones
such that the equal-time basic commutation relations of QED are preserved.     
The advantage of the method is that the calculation of the Green
function is -- similar to the determination of the mode structure
in the standard scheme -- a purely classical problem. The
Green function is essentially determined by the permittivity
of the medium, which is a space-dependent, complex function of
frequency. The configuration of the dielectric bodies is
described by the dependence on space of the permittivity, and 
the effects of dispersion and absorption, respectively, are described
by its real and imaginary parts. It is worth noting
that they are not independent of each other, but they must satisfy 
Kramers--Kronig relations, because of causality (see, e.g., \cite{41,42}). 
The Green function method has been proved correct for radiation  
in 3D bulk material \cite{39,43} and in 1D multilayer structures 
\cite{39,40,43}, and applications to various problems have been studied 
(e.g., ground-state field fluctuations \cite{40,44}, 
photonic wave packets at dielectric barriers \cite{45}, 
nonclassical-light propagation in dispersive and absorbing dielectrics
\cite{46,47}). 

The aim of this paper is to extend the Green function method 
to the quantization of the electromagnetic field in a
dispersive and absorbing 3D inhomogeneous medium. For this purpose,
both the transverse and the longitudinal parts of the electromagnetic
field vectors must be included in the analysis in an unified manner. Relating 
the overall noise current to a bosonic basic field, the
electromagnetic field operators can be expressed -- through
the dyadic Green function of the classical problem -- in terms
of this field, and all the fundamental electromagnetic-field commutation 
relations can be expressed in terms of the Green function. 
Using earlier results of the calculation of the classical 3D Green 
function for multilayer systems \cite{49,50,51,52}, we explicitly
consider an inhomogeneous medium that is composed of two bulk
dielectrics with a common planar interface. 

The paper is organized as follows. 
In Sec.~\ref{sec2} the quantization scheme is developed. 
The Green function expansions of both the electromagnetic
field vectors and the vector and scalar potentials in Coulomb gauge are
given, and the fundamental commutation relations of QED  are studied.
In Sec.~\ref{sec3} the theory is applied to a bulk dielectric, 
and it is shown that the equal-time basic commutation relations of QED 
are preserved. In particular, earlier results for the transverse 
electromagnetic field are recovered. Quantization of the electromagnetic field in
an inhomogeneous medium that consists of two 
bulk dielectrics with a common interface is studied in detail in 
Sec.~\ref{sec4}. Finally, a summary and conclusions are given in 
Sec.~\ref{sec5}. Lengthy mathematical derivations are outlined in appendices.


\section{Quantization scheme}
\label{sec2}
\subsection{Classical Maxwell equations}
\label{sec2a}

Let us start by writing the phenomenological Maxwell equations 
in the frequency domain as
\begin{eqnarray}
\nabla\cdot\underline{\bf B}({\bf r},\omega) & = & 0, 
\label{e1}\\
\nabla\cdot\underline{\bf D}({\bf r},\omega) & = & 0, 
\label{e2}\\
\nabla\times\underline{\bf E}({\bf r},\omega) 
& = & i\omega \,\underline{\bf B}({\bf r},\omega), 
\label{e3}\\
\nabla\times\underline{\bf H}({\bf r},\omega) 
& = & - i\omega \,\underline{\bf D}({\bf r},\omega),
\label{e4}
\end{eqnarray}
where we have assumed that no ``visible'' charges and
currents are embedded in the background medium. The Maxwell equations 
must be supplemented with constitutive 
relations, which for linear dielectric media are usually given by
\begin{eqnarray}
\underline{\bf D}({\bf r},\omega) 
&=& \epsilon_0 \epsilon({\bf r},\omega) \, \underline{\bf E}({\bf r},\omega), 
\label{e5}\\
\underline{\bf B}({\bf r},\omega) &=& \mu_0 \underline{\bf H}({\bf r},\omega). 
\label{e6}
\end{eqnarray}
Here, the (relative) permittivity, which is a complex function of frequency, 
\begin{eqnarray}
\epsilon({\bf r},\omega) 
= \epsilon_{\rm R}({\bf r},\omega) + i\epsilon_{\rm I}({\bf r},\omega),
\label{e7}
\end{eqnarray}
is also allowed to be varying with space in order to model inhomogeneous media.
For causality reasons, the real and imaginary parts of the permittivity,
which are responsible for dispersion and absorption, respectively, 
are uniquely related to each other through Kramers--Kronig relations,
i.e., dispersion and absorption are intimately linked.   
It can be shown that $\epsilon({\bf r},\Omega)$ as function of the 
complex frequency $\Omega$ is analytic and has no zeros in the upper 
complex half-plane, and $\epsilon({\bf r},\Omega)$ $\!\to$ $\!1$
if $|\Omega|$ $\!\to$ $\!\infty$ \cite{41,42}. The fields in the time 
domain are obtained by Fourier transforming the fields in the frequency 
domain, e.g.,  
\begin{eqnarray}
{\bf E} ({\bf r},t) = \int_0^\infty d\omega \, e^{-i\omega t}
\underline{\bf E} ({\bf r},\omega) + {\rm c.c.},
\label{e8}
\end{eqnarray}
and ${\bf B}({\bf r},t)$, ${\bf D}({\bf r},t)$, and ${\bf H}({\bf r},t)$ 
accordingly.

The Maxwell equations (\ref{e1}) -- (\ref{e4}) together with the
constitutive relations (\ref{e5}) and (\ref{e6}) cannot be transferred
to quantum theory by simply regarding the electromagnetic field 
vectors as operator-valued quantities, otherwise the operators would be
damped to zero. This is not surprising, because equations of the
form given here violate, in general, the dissipation--fluctuation 
theorem, which states that damping is always connected with 
additional noise. In other words, even the classical equations are 
equations for the field averages but not equations for the ``naked'' 
fields, and therefore they cannot be used to study the statistics 
of fluctuating fields, such as thermal fields. Hence, transferring the
above given equations to quantum theory can only yield equations for the 
(now quantum-mechanical) expectation values of the fields (which
of course can be damped to zero). The noise that is unavoidably 
associated with absorption can be described by introducing a 
corresponding source term in the Maxwell equations~\cite{40a,52aa}, 
which can be 
thought of as to arise from a noise polarization in the constitutive 
relation between the dielectric displacement vector and the vector of 
the electric field strength (see, e.g., \cite{52aaa}),
\begin{eqnarray}
\underline{\bf D}({\bf r},\omega) 
&=& \epsilon_0 \epsilon({\bf r},\omega)\,\underline{\bf E}({\bf r},\omega) 
+ \underline{\bf P}({\bf r},\omega). 
\label{e13}
\end{eqnarray}
Before specifying the noise source let us first turn to quantum theory. 


\subsection{Quantum Maxwell equations}
\label{sec2b}

The Maxwell equations (\ref{e1}) -- (\ref{e4})
together with the constitutive relations (\ref{e6}) and (\ref{e13})
can be transferred to quantum theory, regarding the electromagnetic
field vectors and the noise polarization field vector as operators:
\begin{equation}
\nabla\cdot \hat{\underline{\bf B}}({\bf r},\omega) = 0, 
\label{e9}
\end{equation}
\begin{equation}
\nabla\cdot  \left[ \epsilon_0 \epsilon({\bf r},\omega)\,
\hat{\underline{\bf E}}({\bf r},\omega) \right] 
= \hat{\underline\rho}({\bf r},\omega), 
\label{e10}
\end{equation}
\begin{equation}
\nabla\times \hat{\underline{\bf E}}({\bf r},\omega) 
= i\omega \,\hat{\underline{\bf B}}({\bf r},\omega), 
\label{e11}
\end{equation}
\begin{equation}
\nabla\times \hat{\underline{\bf B}}({\bf r},\omega)  
= -i\omega \mu_0\epsilon_0 \epsilon({\bf r},\omega) \,
\hat{\underline{\bf E}}({\bf r},\omega) 
+ \mu_0\hat{\underline{\bf j}}({\bf r},\omega).
\label{e12}
\end{equation}
Here, the operator noise charge density $\hat{\underline\rho}$ and 
the operator noise current density $\hat{\underline{\bf j}}$ are
introduced, which are related to the operator noise polarization 
$\hat{\underline{\bf P}}$ as 
\begin{eqnarray}
\hat{\underline\rho}({\bf r},\omega) 
& = & - \nabla\cdot \hat{\underline{\bf P}}({\bf r},\omega), 
\label{u13} \\
\hat{\underline{\bf j}}({\bf r},\omega) 
& = & - i\omega \hat{\underline{\bf P}}({\bf r},\omega).
\label{u13a} 
\end{eqnarray}
It follows from Eqs.~(\ref{u13}) and (\ref{u13a}) that 
$\hat{\underline\rho}$ and $\hat{\underline{\bf j}}$ fulfill the 
equation of continuity,
\begin{eqnarray}
\nabla\cdot \hat{\underline{\bf j}}({\bf r},\omega) 
= i\omega \hat{\underline\rho}({\bf r},\omega). 
\label{e15}
\end{eqnarray}
The electric-field strength operator $\hat{\bf E} ({\bf r})$ (in the
Schr\"{o}\-dinger picture) is defined in terms of the Fourier transform 
$\underline{\hat{\bf E}} ({\bf r},\omega)$ as
\begin{eqnarray}
\hat{\bf E} ({\bf r}) = \int_0^\infty d\omega \, 
\underline{\hat{\bf E}} ({\bf r},\omega) + {\rm H.c.},
\label{e20a}
\end{eqnarray}
and similar relations hold for 
$\hat{\bf B}({\bf r})$, $\hat{\bf D}({\bf r})$, 
and $\hat{\bf H}({\bf r})$.

As already mentioned, the source terms $\hat{\underline\rho}$
and $\hat{\underline{\bf j}}$ are closely related to the noise 
associated with the losses in the medium, which themselves are
described by the imaginary part of the permittivity. Following 
\cite{39,40}, we relate $\hat{\underline{\bf j}}$ to a bosonic vector
field $\hat{\bf f}$ as 
\begin{eqnarray}
  \hat{\underline{\bf j}}({\bf r},\omega) = {\omega \over \mu_0 c^2}
  \sqrt{ {\hbar\over\pi\epsilon_0} \epsilon_{\rm I}({\bf r},\omega) }
  \, \hat{\bf f}({\bf r},\omega),
\label{e17}
\end{eqnarray}
\begin{eqnarray}
  \Bigl[ \hat{f}_i({\bf r},\omega ), \hat{f}^\dagger_j({\bf r}',\omega')
  \Bigr] 
  &=& \delta_{ij}\delta({\bf r}-{\bf r}')\delta(\omega-\omega'),
\label{e18}\\
  \Bigl[ \hat{f}_i ({\bf r},\omega ),\hat{f}_j ({\bf r}',\omega')
  \Bigr]
  &=& 0 = 
  \Bigl[\hat{f}^\dagger_i({\bf r},\omega), 
        \hat{f}^\dagger_j({\bf r}',\omega')\Bigr].
\label{e19}
\end{eqnarray}
Obviously, in the Heisenberg picture the basic operator field evolves
as $\hat{\bf f}({\bf r},\omega,t)$ $\!=$ $\!\hat{\bf f}({\bf r},\omega,t')$
$\!\exp[-i\omega(t$ $\!-$ $\!t')]$, which is governed by the Hamiltonian
\begin{eqnarray}
\hat{H}= \int d^3{\bf r} \int_0^\infty d\omega \, \hbar\omega \, 
\hat{\bf f}^\dagger({\bf r},\omega)\cdot \hat{\bf f}({\bf r},\omega).
\label{e20}
\end{eqnarray}

The system of equations (\ref{e9}) -- (\ref{e12}) together with 
Eqs.~(\ref{e20a}) -- (\ref{e20}) is complete, i.e., 
further equations are not required. 
In particular, all the electromagnetic-field commutation 
relations are uniquely determined from the equations given. 
It should be pointed out that -- in contrast to \cite{39} -- the 
current density $\hat{\underline{\bf j}}$ is not transverse,
because the whole electromagnetic field is considered. Hence,
the vector field $\hat{\bf f}$ introduced here is not 
transverse as well, and the spatial $\delta$ function in
Eq.~(\ref{e18}) is an ordinary $\delta$ function instead of a 
transverse one.


\subsection{Integral representation of 
$\hat{\underline{\bf E}}({\bf r},\omega)$ and
$\hat{\underline{\bf B}}({\bf r},\omega)$}
\label{sec2c}

Equations (\ref{e11}) and (\ref{e12}) imply that  
the electric field $\hat{\underline{\bf E}}({\bf r},\omega)$ obeys 
the partial differential equation
\begin{eqnarray}
\nabla\times\nabla\times \hat{\underline{\bf E}}({\bf r},\omega) 
- {\omega^2\over c^2} \epsilon({\bf r},\omega) 
\hat{\underline{\bf E}}({\bf r},\omega)
= i\omega \mu_0 \hat{\underline{\bf j}}({\bf r},\omega),
\label{e21}
\end{eqnarray}
whose solution can be represented as 
\begin{eqnarray}
\hat{\underline{E}}_i({\bf r},\omega) 
= i\omega\mu_0 \int d^3{\bf s} \, G_{ij}({\bf r}, {\bf s},\omega) 
\,\, \hat{\!\underline{j}}_j({\bf s},\omega),
\label{e22}
\end{eqnarray}
where $\hat{\underline{\bf j}}$ is given by Eq.~(\ref{e17}),
and $G_{ij}({\bf r}, {\bf s},\omega)$ is the dyadic Green function 
(Green tensor) of the classical problem. It satisfies the equation
\begin{eqnarray}
\lefteqn{
\left[ \partial^r_i \partial^r_m - \delta_{im} 
\left(\Delta^r+ {\omega^2\over c^2}\epsilon({\bf r},\omega)\right) \right]
G_{mj}({\bf r}, {\bf s},\omega) 
}
\nonumber \\ && \hspace{35ex}  
= \delta_{ij}\delta({\bf r}-{\bf s}),
\label{e23}
\end{eqnarray}
together with appropriate boundary conditions. In particular, 
it must vanish at infinity. The notation $\partial^r_i$
means $\partial/\partial x_i$, and $\Delta^r=\partial^r_i\partial^r_i$ 
(here and in the following the summation convention is used).
When the electric field $\hat{\underline{\bf E}}({\bf r},\omega)$ is
known, then the magnetic field $\hat{\underline{\bf B}}({\bf r},\omega)$ 
can be obtained as, on recalling Eq.~(\ref{e11}),
\begin{eqnarray}
\hat{\underline{\bf B}}({\bf r},\omega)
= (i\omega)^{-1} \nabla\times\hat{\underline{\bf E}}({\bf r},\omega).
\label{e22a}
\end{eqnarray}
In this way, the electromagnetic field operators can be expressed 
in terms of the bosonic field 
$\hat{\bf f}({\rm r},\omega)$.


\subsection{Commutation relations}
\label{sec2d}

Let us consider the (equal-time) commutation relations between the 
fundamental fields $\hat{\bf E}({\bf r})$ and $\hat{\bf B}({\bf r})$. 
Recalling the definitions of $\hat{\bf E}({\bf r})$ and
$\hat{\bf B}({\bf r})$ [see Eq.~(\ref{e20a})], using Eqs.~(\ref{e22})
and (\ref{e22a}) together with Eq.~(\ref{e17}), and applying the 
commutation relations (\ref{e18}) and (\ref{e19}), we find that
\begin{eqnarray}
\lefteqn{
[\hat{E}_i({\bf r}),\hat{B}_k({\bf r}')]
= {i\hbar \over \pi\epsilon_0}\epsilon_{kmj} \partial^{r'}_m 
\int_0^\infty d\omega 
}
\nonumber \\ && \hspace{2ex} \times
\, {\omega^3\over c^4} 
\int d^3{\bf s} \, \epsilon_{\rm I}({\bf s},\omega)
G_{il}({\bf r}, {\bf s},\omega)  G^*_{jl}({\bf r}', {\bf s},\omega)
 - {\rm c.c.},
\label{e28}
\end{eqnarray}
where $\epsilon_{kmj}$ is the Levi--Civita tensor. 
In order to simplify Eq.~(\ref{e28}), we first note that the relation
\begin{eqnarray}
\lefteqn{
{\omega^2\over c^2} \int d^3{\bf s} \,
\epsilon_{\rm I}({\bf s},\omega)
G_{il}({\bf r}, {\bf s},\omega)G^*_{jl}({\bf r}', {\bf s},\omega)
}
\nonumber \\ && \hspace{8ex}
= {1\over 2i}\left[
G_{ij}({\bf r}, {\bf r}',\omega) - 
G^*_{ij}({\bf r}, {\bf r}',\omega) \right] 
\label{e28a}
\end{eqnarray}
is valid (see App.~\ref{secapp1}). Further, from Eq.~(\ref{e23}) 
and the relation $\epsilon^*({\bf r},\omega)$ $\!=$ 
$\!\epsilon({\bf r},-\omega)$ 
it follows that
\begin{eqnarray}
G^*_{ij}({\bf r}, {\bf r}',\omega) = G_{ij}({\bf r}, {\bf r}',-\omega).
\label{e28aa}
\end{eqnarray}
Combining Eqs.~(\ref{e28}) -- (\ref{e28aa}), we derive 
\begin{eqnarray}
\lefteqn{
\Bigl[
  \hat{E}_i({\bf r}),\hat{B}_k({\bf r}') \Bigr] 
}
\nonumber \\ && \hspace{5ex}  
= {\hbar\over \pi\epsilon_0} \epsilon_{kmj} \partial^{r'}_m 
\int_{-\infty}^\infty d\omega \, {\omega\over c^2} 
G_{ij}({\bf r}, {\bf r}',\omega).
\label{e29}
\end{eqnarray} 
Similarly, we find that
\begin{eqnarray}
\Bigl[
  \hat{E}_i({\bf r}),\hat{E}_k({\bf r}')\Bigr] = 0 = 
\Bigl[\hat{B}_i({\bf r}),\hat{B}_k({\bf r}')\Bigr],
\label{e30}
\end{eqnarray}
which is in full agreement with QED (see, e.g., \cite{52a}). Equation (\ref{e29}) reveals that the 
commutator between the electric and magnetic fields can be expressed in 
terms of a single frequency integral of the Green function multiplied by 
the frequency. In order to calculate this integral, knowledge of the Green 
function is required. 
Note that a single pole at $\omega$ $\!=$ $\!0$ has to be treated
as principal value.


\subsection{Vector potential and scalar potential}
\label{sec2e}
\subsubsection{Potential equations and integral representations}
\label{sec2e1}

It is often 
necessary
to use electromagnetic potentials.
In the frequency domain, the vector and scalar
potentials $\hat{\underline{\bf A}}$ and $\hat{\underline\varphi}$,
respectively, are related to the fields as 
\begin{eqnarray}
\hat{\underline{\bf B}}({\bf r},\omega)  
&=& \nabla\times\hat{\underline{\bf A}}({\bf r},\omega) ,
\label{e32}\\
\hat{\underline{\bf E}}({\bf r},\omega)  
&=& i\omega \hat{\underline{\bf A}}({\bf r},\omega)  
- \nabla\hat{\underline\varphi}({\bf r},\omega) .
\label{e33} 
\end{eqnarray}
Substituting in Eqs.~(\ref{e10}) and ~(\ref{e12}) for the fields
the potentials according to Eqs.~(\ref{e32}) and (\ref{e33}) then yields
\begin{eqnarray}
\lefteqn{
\nabla\cdot[\epsilon({\bf r},\omega)
\nabla\hat{\underline\varphi}({\bf r},\omega)] 
}
\nonumber \\ && \hspace{10ex}
= -{\hat\rho\over \epsilon_0} +
i\omega \nabla\cdot[\epsilon({\bf r},\omega)
\hat{\underline{\bf A}}({\bf r},\omega)],
\label{e34}
\end{eqnarray}
\begin{eqnarray}
\lefteqn{
\nabla\times\nabla\times\hat{\underline{\bf A}}({\bf r},\omega) 
  - {\omega^2\over c^2} \epsilon({\bf r},\omega) 
  \hat{\underline{\bf A}}({\bf r},\omega)
}
\nonumber \\ && \hspace{10ex}  
=\mu_0 \hat{\underline{\bf j}}({\bf r},\omega) 
+ {i\omega\over c^2} \epsilon({\bf r},\omega)
\nabla\hat{\underline\varphi}({\bf r},\omega).
\label{e35} 
\end{eqnarray}

In Coulomb gauge,
\begin{eqnarray}
\nabla\cdot\hat{\underline{\bf A}}({\bf r},\omega)=0,
\label{e36} 
\end{eqnarray}
Eq.~(\ref{e33}) corresponds -- in the sense of the Helmholtz theorem --
to a unique decomposition of the electric field $\hat{\underline{\bf E}}$ 
into a transverse part $i\omega\hat{\underline{\bf A}}$ and a 
longitudinal part $- \nabla\hat{\underline\varphi}$. 
Hence we may write
\begin{eqnarray}
\hat{\underline{A}}_i({\bf r},\omega) 
= (i\omega)^{-1} \int d^3{\bf s} \,
\delta^\perp_{ij}({\bf r}-{\bf s}) \hat{\underline{E}}_j({\bf s},\omega),
\label{e37}\\
\partial^{r}_{i}\hat{\underline\varphi}({\bf r},\omega)
= - \int d^3{\bf s} \,
\delta^\parallel_{ij}({\bf r}-{\bf s}) \hat{\underline{E}}_j({\bf s},\omega),
\label{e38}
\end{eqnarray}
where $\delta^\perp({\bf r})$ and $\delta^\parallel({\bf r})$,
respectively, are the transverse and longitudinal $\delta$ functions
(see, e.g., \cite{2}),
\begin{eqnarray}
\delta^\perp_{ij}({\bf r}) &=& \delta_{ij}\delta({\bf r})
+ \partial^r_i\partial^r_j (4\pi r)^{-1},
\label{e38a}\\
\delta^\parallel_{ij}({\bf r}) &=&
- \partial^r_i\partial^r_j (4\pi r)^{-1}.
\label{e38b}
\end{eqnarray}
We insert Eq.~(\ref{e22}) into Eq.~(\ref{e37}) and obtain the 
following integral representation of the vector potential: 
\begin{eqnarray}
\hat{\underline{A}}_i({\bf r},\omega) = \mu_0
\int d^3{\bf s} \, G^{\!\perp}_{im}({\bf r}, {\bf s},\omega)
\,\hat{\!\underline{j}}_m ({\bf s},\omega),
\label{e39}
\end{eqnarray}
where
\begin{eqnarray}
G^{\!\perp}_{im}({\bf r}, {\bf s},\omega) = \int d^3{\bf s}' \,
\delta^\perp_{ij}({\bf r}-{\bf s}') G_{jm}({\bf s}', {\bf s},\omega)
\label{e39a}
\end{eqnarray}
is the (from the left) one-sided transverse Green tensor.
Next we substitute in Eq.~(\ref{e38}) for the longitudinal $\delta$
function the expression (\ref{e38b}) and find that
\begin{eqnarray}
\hat{\underline\varphi}({\bf r},\omega) 
= \partial^r_j \int d^3{\bf s} \,
\frac{\hat{\underline{E}}_j({\bf s},\omega) }{4\pi |{\bf r}-{\bf s}|} \,.
\label{e40} 
\end{eqnarray}
The integral representation of the scalar potential can then be found,
substituting in Eq.~(\ref{e40}) for the electric field the
integral representation (\ref{e22}).  


\subsubsection{Commutation relations}
\label{sec2e2}

It is well known that $\hat{\bf A}({\bf r})$ and 
$\epsilon_0\dot{\hat{\bf A}}({\bf r})$ are canonically conjugated
field variables. In order to calculate the commutation relation between 
them and the scalar potential $\hat{\varphi}({\bf r})$, 
we first note that $\hat{\bf A}({\bf r})$,
$\dot{\hat{\bf A}}({\bf r})$, and $\hat{\varphi}({\bf r})$ are given 
by integrals of the type (\ref{e20a}), but with 
$\hat{\underline{\bf A}}({\bf r},\omega)$,
$-i\omega\hat{\underline{\bf A}}({\bf r},\omega)$, 
and $\hat{\underline\varphi}({\bf r},\omega)$, respectively,
in place of $\hat{\underline{\bf E}}({\bf r},\omega)$.
Using Eq.~(\ref{e39}) and following the lines outlined for calculating
the commutation relations (\ref{e29}) and (\ref{e30}), we obtain 
\begin{eqnarray}
\Bigl[
  \hat{A}_i({\bf r}),\,\dot{\!\hat{A}}_j({\bf r}') \Bigr] = 
{\hbar\over \pi\epsilon_0}  
\int_{-\infty}^\infty d\omega \, {\omega\over c^2} 
G^{\perp\!\!\perp}_{ij}({\bf r}, {\bf r}',\omega)
\label{e45}
\end{eqnarray}
and
\begin{eqnarray}
\Bigl[
  \hat{A}_i({\bf r}),\hat{A}_j({\bf r}')
\Bigr] = 0 =
\Bigl[
  \,\dot{\!\hat{A}}_i({\bf r}),\,\dot{\!\hat{A}}_j({\bf r}')
\Bigr],
\label{e46}
\end{eqnarray}
where 
\begin{eqnarray}
\lefteqn{
G^{\perp\!\!\perp}_{ij}({\bf r}, {\bf r}',\omega) 
}
\nonumber \\ && \hspace{1ex}
= \int \! d^3{\bf s} \int \! d^3{\bf s}' \,
\delta^\perp_{im}({\bf r}-{\bf s}) G_{mn}({\bf s}, {\bf s}',\omega)
\delta^\perp_{nj}({\bf s}'-{\bf r}')
\label{e45a}
\end{eqnarray}
is the two-sided transverse Green tensor.
Similarly, the commutation relation between the vector potential and 
the scalar potential can be given by
\begin{eqnarray}
\lefteqn{
\Bigl[
\hat\varphi({\bf r}),\hat{A}_j({\bf r}')\Bigr] 
}
\nonumber \\ && \hspace{5ex}
= {\hbar\over \pi\epsilon_0} \partial^r_m \int d^3{\bf s} 
\int_{-\infty}^\infty d\omega \, {\omega\over c^2}
\frac{G^{\hspace{1ex}\perp}_{mj}({\bf s}, {\bf r}',\omega)}
{4\pi |{\bf r}-{\bf s}|}\,,
\label{e47}
\end{eqnarray}
where
\begin{eqnarray}
G^{\hspace{1ex}\perp}_{mj}({\bf s}, {\bf r}',\omega)
= \int d^3{\bf s}' \, G_{mn}({\bf s}, {\bf s}',\omega) 
\delta^\perp_{nj}({\bf s}'-{\bf r}')
\label{e47a}
\end{eqnarray}
is the (from the right) one-sided transverse Green tensor,
and finally 
\begin{eqnarray}
\Bigl[  \hat\varphi({\bf r}), \hat\varphi({\bf r}') \Bigr] = 0 = 
\Bigl[
\hat\varphi({\bf r}),\,\dot{\!\hat{A}}_i({\bf r}')\Bigr].
\label{e47aa}
\end{eqnarray}
In order to further calculate the commutators (\ref{e45}) and
(\ref{e47}), the Green function multiplied by the frequency
must be integrated over frequency, which is quite similar to the
commutation relation (\ref{e29}).


\section{Homogeneous dielectrics}
\label{sec3}


Let us first consider the electromagnetic field in an absorbing 
bulk material such that the permittivity can be assumed to be
independent of space: $\epsilon({\bf r},\omega)$ $\!=$ 
$\!\epsilon(\omega)$ $\!=$ $\!\epsilon_{\rm R}(\omega)$ 
$\!+$ $\!i\epsilon_{\rm I}(\omega)$ for all ${\bf r}$. 
In this case, the solution of Eq.~(\ref{e23}) which satisfies
the boundary conditions at infinity is \cite{52}
\begin{eqnarray}
\lefteqn{
G_{ij}({\bf r}, {\bf r}',\omega)
}
\nonumber \\ && \hspace{5ex}
=\left[ \partial^r_i\partial^{r}_j + \delta_{ij} q^2(\omega) \right]
q^{-2}(\omega) \, g(|{\bf r}-{\bf r}'|,\omega),
\label{e50}
\end{eqnarray}
where the notation $q^2(\omega)$ $\!=$ $\!(\omega^2/c^2)\epsilon(\omega)$ is
used, and
\begin{eqnarray}
g(|{\bf r}-{\bf r}'|,\omega) = 
{e^{ iq(\omega)|{\bf r}-{\bf r}'|} \over 4\pi |{\bf r}-{\bf r}'|}\,.
\label{e52}
\end{eqnarray}
Substituting in Eq.~(\ref{e22}) for the Green function the function
given in Eq.~(\ref{e50}) and integrating over frequency yields, 
together with Eq.~(\ref{e17}), the Green function expansion of the 
operator of the electric field strength of the electromagnetic
field in a dispersive and absorbing bulk dielectric. 
The Green function expansions of the magnetic field and the
potentials can be obtained accordingly.  


\subsection{Commutation relations} 
\label{sec3a}

We insert the Green function (\ref{e50}) into Eq.~(\ref{e29}) and find that
\begin{eqnarray}
\lefteqn{
\Bigl[
  \hat{E}_i({\bf r}),\hat{B}_k({\bf r}')\Bigr] 
}
\nonumber \\ && \hspace{5ex}  
= 
{\hbar\over \pi\epsilon_0} \epsilon_{kmi} \partial^{r'}_m 
\int_{-\infty}^\infty d\omega \, {\omega\over c^2} 
\, g(|{\bf r}-{\bf r}'|,\omega)
\label{e53}
\end{eqnarray}
[note that $\epsilon_{kmj}\partial^{r'}_m\partial^{r'}_j$ $\!(...)$
$\!=$ $\!0$, because of the antisymmetry of the Levi--Civita tensor].
We now substitute in Eq.~(\ref{e53}) for $g(|{\bf r}$ $\!-$ 
$\!{\bf r}'|,\omega)$ the Fourier expansion
\begin{eqnarray}
g(|{\bf r}-{\bf r}'|,\omega) = \int {d^3{\bf k}\over(2\pi)^3}  
{ e^{ i{\bf k}\cdot ({\bf r}-{\bf r}')} \over k^2-q^2(\omega)} \, ,
\label{e54}
\end{eqnarray}
which enables us to calculate the $\omega$ integral by means of contour
integral techniques. Recalling the properties of $\epsilon(\Omega)$ as 
a function of the complex frequency $\Omega$, we obtain (cf. \cite{39,40}) 
\begin{eqnarray}
\int_{-\infty}^{\infty} d\omega \, 
{\omega\over c^2} {1\over k^2-q^2(\omega)} = i\pi \,.
\label{e55}
\end{eqnarray}
Since the result is independent of $k$, the $k$ integral simply
gives a $\delta$ function, and we eventually arrive at
\begin{eqnarray}
\Bigl[ \hat{E}_i({\bf r}),\hat{B}_k({\bf r}') \Bigr] 
=
-{i\hbar\over \epsilon_0} \epsilon_{ikm} \partial^{r}_m 
\delta({\bf r}-{\bf r}').
\label{e56}
\end{eqnarray}
   From Eqs.~(\ref{e56}) and (\ref{e30})
we see that the quantization scheme yields exactly the
equal-time electromagnetic-field commutation relations that are
well established in QED. 

In order to calculate the commutator $[\hat{A}_i({\bf r}),
\,\dot{\!\hat{A}}_j({\bf r}')]$, we combine Eqs.~(\ref{e45}), (\ref{e45a}),
and (\ref{e50}). After a partial integration the derivatives in 
Eq.~(\ref{e50}) give rise to a divergence of the transverse $\delta$ 
function in Eq.~(\ref{e45}) which
vanishes, $\partial^r_i \delta^\perp_{ij}({\bf r})$ $\!=$ $\!0$, so that
\begin{eqnarray}
\lefteqn{
\Bigl[
  \hat{A}_i({\bf r}),\,\dot{\!\hat{A}}_j({\bf r}')\Bigr] 
=
{\hbar\over \pi\epsilon_0}
\int d^3{\bf s} \int d^3{\bf s}' \,
\delta^\perp_{im}({\bf r}-{\bf s})
}
\nonumber \\ && \hspace{5ex}
\times\,\int_{-\infty}^\infty d\omega {\omega\over c^2} \,
g(|{\bf s}-{\bf s}'|,\omega)
\,\delta^\perp_{mj}({\bf s}'-{\bf r}').
\label{e58}
\end{eqnarray}
We again make use of Eqs.~(\ref{e54}) and (\ref{e55}), and 
after some straightforward calculation we obtain
\begin{eqnarray}
\Bigl[
  \hat{A}_i({\bf r}),\,\dot{\!\hat{A}}_j({\bf r}')\Bigr] 
=
{i\hbar\over \epsilon_0} \,\delta^\perp_{ij}({\bf r}-{\bf r}').
\label{e58a}
\end{eqnarray}
Quite similarly, it can be proved that
\begin{eqnarray}
\Bigl[\hat\varphi({\bf r}),\hat{A}_j({\bf r}')\Bigr] 
= 0 .
\label{e59}
\end{eqnarray}
Again, the correct QED equal-time commutation relations are 
preserved. Note that from Eqs.~(\ref{e33}), (\ref{e58a}), and (\ref{e59})  
it follows that
\begin{eqnarray}
\Bigl[ \hat{A}_i({\bf r}),\hat{E}_j({\bf r}') \Bigr] 
= -{i\hbar\over\epsilon_0} \, \delta^\perp_{ij}({\bf r}-{\bf r}'). 
\label{e59a}
\end{eqnarray}


\subsection{Relation to earlier work}
\label{sec3b}

To make contact with earlier work, we first note that, 
according to the Helmholtz theorem, the noise current
$\hat{\underline{\bf j}}$ can be decomposed in a unique way into a 
transverse and a longitudinal part,
\begin{eqnarray}
\hat{\underline{\bf j}}({\bf r},\omega) = 
\hat{\underline{\bf j}}^\perp({\bf r},\omega)
+ \hat{\underline{\bf j}}^\parallel({\bf r},\omega),
\label{e60}
\end{eqnarray}
where
\begin{eqnarray}
\,\hat{\!\underline j}^{\perp(\parallel)}_i({\bf r},\omega) = \int d^3{\bf s}
\,\delta^{\perp(\parallel)}_{ij}({\bf r}-{\bf s}) \,
\,\hat{\!\underline j}_j ({\bf s},\omega) .
\label{e61}
\end{eqnarray}
Since the permittivity is assumed to be independent of space, 
the decomposition of $\hat{\underline{\bf j}}$ corresponds
to a decomposition of the bosonic (basic) field $\hat{\bf f}$ into a 
transverse and a longitudinal part. We use Eq.~(\ref{e17}) and 
can relate $\hat{\underline{\bf j}}^{\perp(\parallel)}$
to $\hat{\bf f}^{\perp(\parallel)}$ as
\begin{eqnarray}
\hat{\underline{\bf j}}^{\perp(\parallel)}({\bf r},\omega) 
= {\omega\over \mu_0 c^2}
\sqrt{ {\hbar\over\pi\epsilon_0} \epsilon_{\rm I}(\omega) } \,
\hat{\bf f}^{\perp(\parallel)}({\bf r},\omega),
\label{e62}
\end{eqnarray}
where 
\begin{eqnarray}
\hat{f}^{\perp(\parallel)}_i({\bf r},\omega) 
= \int d^3{\bf s} \; \delta^{\perp(\parallel)}_{ij}({\bf r}-{\bf s}) \,
\hat{f}_j ({\bf s},\omega).
\label{e63}
\end{eqnarray}
The commutation relations (\ref{e18}) and (\ref{e19}) obviously
imply that 
\begin{eqnarray}
\lefteqn{
\left[ \hat{f}^{\perp(\parallel)}_i({\bf r},\omega), 
     \Bigl( \hat{f}^{\perp(\parallel)}_j({\bf r}',\omega') \Bigr)^\dagger 
\right] 
}
\nonumber \\ && \hspace{10ex}
= \delta^{\perp(\parallel)}_{ij}({\bf r}-{\bf r}')
\,\delta(\omega-\omega'),
\label{e64a}
\end{eqnarray}
\begin{eqnarray}
\lefteqn{
\left[ \hat{f}^{\perp(\parallel)}_i({\bf r},\omega), 
       \hat{f}^{\perp(\parallel)}_j({\bf r}',\omega') 
\right] 
}
\nonumber \\ && \hspace{5ex}
= 0 =
\left[ \Bigl(\hat{f}^{\perp(\parallel)}_i({\bf r},\omega) \Bigr)^\dagger, 
       \Bigl(\hat{f}^{\perp(\parallel)}_j({\bf r}',\omega') \Bigr)^\dagger 
\right], 
\label{e64}
\end{eqnarray}
and
\begin{eqnarray}
\left[ \hat{f}^\perp_i({\bf r},\omega), 
     \Bigl( \hat{f}^\parallel_j({\bf r}',\omega') \Bigr)^\dagger \right]= 0.
\label{e65}
\end{eqnarray}
Obviously, the motion of the transverse and longitudinal degrees of freedom 
are governed by their own Hamiltonians, as can be readily seen
from the Hamiltonian (\ref{e20}), which can be rewritten as
\begin{eqnarray}
\hat{H} &=& \hat{H}^\perp + \hat{H}^\parallel,
\label{e66}
\end{eqnarray}
with
\begin{eqnarray}
\lefteqn{
\hat{H}^{\perp(\parallel)}
}
\nonumber \\ && \hspace{2ex}
= \int d^3{\bf r} \int_0^\infty d\omega \, \hbar\omega 
\left[ \hat{\bf f}^{\perp(\parallel)} ({\bf r},\omega) \right]^\dagger
\cdot \hat{\bf f}^{\perp(\parallel)}({\bf r},\omega).
\label{e66a}
\end{eqnarray}

  From Eqs.~(\ref{e39}), (\ref{e39a}), and (\ref{e50}) it can be
seen that, after partial integration, the derivatives in Eq.~(\ref{e50})
do not contribute to the vector potential, because of the vanishing
divergence of the transverse $\delta$ function.
Therefore we may write    
\begin{eqnarray}
\lefteqn{
\hat{\underline A}_i({\bf r},\omega) 
= \mu_0 \int d^3{\bf s} 
}
\nonumber \\ && \hspace{2ex} \times
\int d^3{\bf s}' \,
\delta^\perp_{ij}({\bf r}-{\bf s}) 
\,g(|{\bf s}-{\bf s}'|,\omega) 
\,\,\hat{\!\underline j}_j ({\bf s}',\omega),
\label{e67}
\end{eqnarray}
or, changing the variable ${\bf s}$ as ${\bf r}'$ $\!=$ 
$\!{\bf r}$ $\!-$ $\!{\bf s}$ $\!+$ $\!{\bf s}'$,
\begin{eqnarray}
\hat{\underline A}_i({\bf r},\omega) 
=
\mu_0 \int d^3{\bf r}' \, 
g(|{\bf r}-{\bf r}'|,\omega) \,
\,\hat{\!\underline j}^\perp_i({\bf r}',\omega),
\label{e68}
\end{eqnarray}
which is nothing but the representation of the transverse vector 
potential given in \cite{39} (if we identify 
$\mu_0\hat{\underline{\bf j}}^\perp$ 
with $\underline{\hat{\bf j}}_n$ in \cite{39}).
Similarly, from Eqs.~(\ref{e22}) and (\ref{e40}) and the Green 
function (\ref{e50}) [together with Eq.~(\ref{e52})] it can be derived that
\begin{eqnarray}
\underline{\hat\varphi}({\bf r},\omega) = {1\over 4\pi\epsilon_0\epsilon(\omega)}
\int d^3{\bf s} \, {\underline{\hat{\rho}}({\bf s},\omega) 
\over |{\bf r}-{\bf s}|}\,,
\label{e69}
\end{eqnarray}
where 
\begin{eqnarray}
\underline{\hat{\rho}}({\bf r},\omega) 
& = & (i \omega)^{-1}
\nabla\cdot \underline{\hat{\bf j}}^\parallel({\bf r},\omega)
\nonumber\\ 
& = &{1 \over i\mu_0 c^2} 
\sqrt{ {\hbar\over\pi\epsilon_0} \epsilon_{\rm I}( \omega) } 
  \, \nabla\cdot \hat{\bf f}^{\parallel}({\bf r},\omega) 
\label{e70}
\end{eqnarray}
[cf. Eq.~(\ref{e15})].
Note that in the case under study from the commutation relations
(\ref{e65}) and Eqs.~(\ref{e68}) and (\ref{e69}) [together with
Eq.~(\ref{e70})] it is immediately seen that vector and scalar 
potentials are commuting quantities, i.e., the commutation relation
(\ref{e59}) is fulfilled.


\section{Dielectric Interface}
\label{sec4}
\subsection{The Green function}
\label{sec4a}

The determination of the dyadic Green function for
three-dimensional configurations of dielectric bodies is a very
involved problem in general, and only for rather simple configurations
the Green function has been calculated explicitly. 
Such a configuration, which can be thought of as being the basic
element of multilayer dielectric structures, is composed of
two infinite half-spaces ($V_1$ and $V_2$) with a common planar 
interface such that
\begin{equation}
\epsilon({\bf r},\omega) = 
\left\{
\begin{array}{l} 
\epsilon_1(\omega) \quad {\rm if} \quad {\bf r}\in V_1, 
\quad {\rm i.e.,} \quad z<0,\\ 
\epsilon_2(\omega)\quad {\rm if} \quad {\bf r}\in V_2,
\quad {\rm i.e.,} \quad z>0. 
\end{array} 
\right. 
\label{e73}
\end{equation}
Following \cite{39,45,52}, we write the solution of 
Eq.~(\ref{e23}) in the form
\begin{eqnarray}
\lefteqn{
G_{ij}({\bf r}, {\bf r}',\omega) 
}
\nonumber \\ && \hspace{2ex}
= 
\left\{ \begin{array}{l}
\!G_{ij}^\alpha({\bf r}, {\bf r}',\omega) 
\!+\! R_{ij}^\alpha({\bf r}, {\bf r}',\omega)
\quad {\rm if} \quad {\bf r}, {\bf r}' \!\in\! V_\alpha, \\[.5ex]
\!T_{ij}^{\alpha\alpha'}({\bf r}, {\bf r}',\omega)
\quad {\rm if} \quad {\bf r}\!\in \!V_{\alpha}, \ {\bf r}'\! \in \!V_{\alpha'}     
\, (\alpha\neq\alpha')
\end{array} \right.
\label{e74}
\end{eqnarray}
($\alpha, \alpha'$ $\!=$ $\!1,2$),
where $G_{ij}^\alpha({\bf r}, {\bf r}',\omega)$ is the Green function 
(\ref{e50}) [together with Eq.~(\ref{e52})] for the bulk material 
with $\epsilon_\alpha(\omega)$,
\begin{eqnarray}
\lefteqn{
G_{ij}^\alpha({\bf r}, {\bf r}',\omega)
}
\nonumber \\ && \hspace{2ex}
=\left[ \partial^r_i\partial^{r}_j + \delta_{ij} q_{\alpha}^{2}(\omega) \right]
q_{\alpha}^{-2}(\omega) \, g^\alpha(|{\bf r}-{\bf r}'|,\omega),
\label{e75}
\end{eqnarray}
with
\begin{eqnarray}
g^\alpha(|{\bf r}-{\bf r}'|,\omega) = 
{ e^{ iq_\alpha(\omega)|{\bf r}-{\bf r}'| } \over 4\pi|{\bf r}-{\bf r}'| }
\label{e77}
\end{eqnarray}
[$q^{2}_{\alpha}(\omega)$ $\! =$ $\!(\omega^2/c^2)\epsilon_\alpha(\omega)$].
The functions $R^\alpha_{ij}({\bf r}, {\bf r}',\omega)$ and 
$T^{\alpha\alpha'}_{ij}({\bf r}, {\bf r}',\omega)$ describe the 
effects of reflection and transmission at the interface and obey the
homogeneous equations
\begin{equation}
\left[ \partial^r_i \partial^r_m 
- \delta_{im} \left(\Delta^r+ q^2_\alpha(\omega) \right) \right]
R^\alpha_{mj}({\bf r}, {\bf r}',\omega) = 0
\label{e78}
\end{equation}
(${\bf r}, {\bf r}'$ $\! \in$ $\! V_\alpha$) and
\begin{equation}
\left[ \partial^r_i \partial^r_m 
- \delta_{im} \left(\Delta^r+ q^2_\alpha(\omega) \right) \right]
T^{\alpha\alpha'}_{mj}({\bf r}, {\bf r}',\omega) = 0
\label{e79}
\end{equation}
(${\bf r}$ $\!\in$ $\! V_{\alpha}$, ${\bf r}'$ $\! \in$ 
$\! V_{\alpha'}$, $\alpha$ $\!\neq$ $\!\alpha'$)
together with the boundary conditions such that they vanish at infinity
and the tangential components of the $\hat{\bf E}$ and the $\hat{\bf H}$ 
field are continuous at the surface of discontinuity.
In order to determine the functions 
$R^\alpha_{ij}({\bf r}, {\bf r}',\omega)$ and 
$T^{\alpha\alpha'}_{ij}({\bf r}, {\bf r}',\omega)$,
it is convenient to take advantage of the translational invariance of 
the system in the $(x,y)$ plane, which enables us to expand the
Green function (and $G_{ij}^{\alpha}$, $R_{ij}^{\alpha}$, and 
$T_{ij}^{\alpha \alpha'}$) 
as follows:
\begin{eqnarray}
G_{ij}({\bf r}, {\bf r}',\omega) = \int {d^2{\bf k}_\parallel\over (2\pi)^2} \,
G_{ij}({\bf k}_\parallel,\omega;z,z') 
e^{ i{\bf k}_\parallel \cdot ({\bf r}_\parallel-{\bf r}_\parallel') },
\label{e80}
\end{eqnarray}
where ${\bf k}_\parallel$ $\!=$ $\!(k_x,k_y,0)$ and ${\bf r}_\parallel$ 
$\!=$ $\!(x,y,0)$ are two-di\-men\-sional vectors in the $(x,y)$ plane.
Using the general formulas for multilayer structures given in \cite{52} 
(see also \cite{49}), after some manipulations we obtain, 
for the reflection functions $R^\alpha_{ij}$ $\!\equiv$
$\!R^\alpha_{ij}({\bf k}_\parallel,\omega;z,z')$,
\begin{eqnarray}
\lefteqn{
R^\alpha_{xx} 
= {i\over 2\beta_\alpha} e^{i\beta_\alpha(|z|+|z'|)}
}
\nonumber \\ && \hspace{10ex} \times
\left[ 
  {r^p_{\alpha\alpha'} \over q_\alpha^2} 
  \left( -\beta_\alpha^2 {k_x^2\over k_\parallel^2} \right)
  + r^s_{\alpha\alpha'}{k_y^2\over k_\parallel^2}
\right],
\label{e81}
\end{eqnarray}
\begin{eqnarray}
\lefteqn{
R^\alpha_{xy} 
= {i\over 2\beta_\alpha} e^{i\beta_\alpha(|z|+|z'|)}
}
\nonumber \\ && \hspace{10ex} \times
\left[ 
  {r^p_{\alpha\alpha'} \over q_\alpha^2} 
  \left( -\beta_\alpha^2 {k_xk_y\over k_\parallel^2} \right)
  - r^s_{\alpha\alpha'}{k_xk_y\over k_\parallel^2}
\right],
\label{e82}
\end{eqnarray}
\begin{equation}
R^\alpha_{xz} 
= {i\over 2\beta_\alpha} e^{i\beta_\alpha(|z|+|z'|)}
  {r^p_{\alpha\alpha'} \over q_\alpha^2}
[ -{\rm sign}(z') \beta_\alpha k_x ],
\label{e83}
\end{equation}
\begin{equation}
R^\alpha_{yx}=R^\alpha_{xy},
\label{e84a}
\end{equation}
\begin{equation} 
R^\alpha_{yy}=R^\alpha_{xx}(k_x\leftrightarrow k_y), \quad 
R^\alpha_{yz}=R^\alpha_{xz}(k_x\leftrightarrow k_y),
\label{e84}
\end{equation}
\begin{equation}
R^\alpha_{zx}=-R^\alpha_{xz}, \quad
R^\alpha_{zy}=-R^\alpha_{yz}, 
\label{e85}
\end{equation}
\begin{equation}
R^\alpha_{zz} 
= {i\over 2\beta_\alpha} e^{i\beta_\alpha(|z|+|z'|)}
  {r^p_{\alpha\alpha'} \over q_\alpha^2} k_\parallel^2,
\label{e86}
\end{equation}
and accordingly for the transmission functions 
$T^{\alpha\alpha'}_{ij}$ $\!\equiv$
$\!T^{\alpha\alpha'}_{ij}({\bf k}_\parallel,\omega;z,z')$,
\begin{eqnarray}
\lefteqn{
T^{\alpha\alpha'}_{xx} 
= {i\over 2\beta_\alpha} e^{i\beta_{\alpha}|z|+i\beta_{\alpha'}|z'|}
}
\nonumber \\ && \hspace{10ex}
\times\,\left( 
  {t^p_{\alpha\alpha'} \over q_\alpha q_{\alpha'}} 
  \beta_\alpha \beta_{\alpha'}{k_x^2\over k_\parallel^2} 
  + t^s_{\alpha\alpha'}{k_y^2\over k_\parallel^2}
\right),
\label{e87}
\end{eqnarray}
\begin{eqnarray}
\lefteqn{
T^{\alpha\alpha'}_{xy} 
= {i\over 2\beta_\alpha} e^{i\beta_{\alpha}|z|+i\beta_{\alpha'}|z'|}
}
\nonumber \\ && \hspace{10ex}
\times\,\left( 
  {t^p_{\alpha\alpha'} \over q_\alpha q_{\alpha'}} 
  \beta_\alpha \beta_{\alpha'}{k_xk_y\over k_\parallel^2} 
  - t^s_{\alpha\alpha'}{k_xk_y\over k_\parallel^2}
\right),
\label{e88}
\end{eqnarray}
\begin{equation}
T^{\alpha\alpha'}_{xz} 
= {i\over 2\beta_\alpha} e^{i\beta_{\alpha}|z|+i\beta_{\alpha'}|z'|}
  {t^p_{\alpha\alpha'} \over q_\alpha q_{\alpha'}} 
  {\rm sign}(z') \beta_{\alpha} k_x,
\label{e89}
\end{equation}
\begin{equation}
T^{\alpha\alpha'}_{yx} = T^{\alpha\alpha'}_{xy}, 
\label{e89a}
\end{equation}
\begin{equation}
T^{\alpha\alpha'}_{yy} 
   = T^{\alpha\alpha'}_{xx}(k_x\leftrightarrow k_y), \quad
T^{\alpha\alpha'}_{yz} 
   = T^{\alpha\alpha'}_{xz}(k_x\leftrightarrow k_y), 
\label{e90}
\end{equation}
\begin{equation}
T^{\alpha\alpha'}_{zx} 
= {i\over 2\beta_\alpha} e^{i\beta_{\alpha}|z|+i\beta_{\alpha'}|z'|}
  {t^p_{\alpha\alpha'} \over q_\alpha q_{\alpha'}} 
  {\rm sign}(z') \beta_{\alpha'} k_x,
\label{e91}
\end{equation}
\begin{equation}
T^{\alpha\alpha'}_{zy} = 
  T^{\alpha\alpha'}_{zx}(k_x\leftrightarrow k_y),
\label{e92}
\end{equation}
\begin{equation}
T^{\alpha\alpha'}_{zz} 
= {i\over 2\beta_\alpha} e^{i\beta_{\alpha}|z|+i\beta_{\alpha'}|z'|}
  {t^p_{\alpha\alpha'} \over q_\alpha q_{\alpha'}} k_\parallel^2,
\label{e93}
\end{equation}
where $\alpha'=1(2)$ for $\alpha=2(1)$, and
\begin{displaymath}
q_\alpha \equiv q_\alpha(\omega), 
\end{displaymath}
\begin{equation}
\beta_\alpha \equiv \beta_\alpha(\omega)
= \sqrt{ q_\alpha^2(\omega) - k_\parallel^2 }, 
\quad
{\rm Re}\,\beta_\alpha \geq 0, \  {\rm Im}\,\beta_\alpha \geq 0,
\label{e94}
\end{equation}
with $r^q_{\alpha\alpha'}$ $\!\equiv$ $r^q_{\alpha\alpha'}(\omega)$
and $t^q_{\alpha\alpha'}$ $\!\equiv$ $t^q_{\alpha\alpha'}(\omega)$,
$q$ $\!=$ $\!p,s$, being the generalized reflection and transmission 
coefficients for the $p$- and $s$-polarized components of the 
electromagnetic field, which are defined by
\begin{equation}
r^q_{\alpha\alpha'} = 
{ \beta_\alpha - \gamma^q_{\alpha\alpha'}\beta_{\alpha'}
  \over
  \beta_\alpha + \gamma^q_{\alpha\alpha'}\beta_{\alpha'} }
  = - r^q_{\alpha'\alpha}, 
\label{e94a}  
\end{equation}  
\begin{equation}
\gamma^p_{\alpha\alpha'} = {\epsilon_\alpha\over \epsilon_{\alpha'}}, \quad
\gamma^s_{\alpha\alpha'} = 1,
\label{e95}
\end{equation}
\begin{equation}
t^q_{\alpha\alpha'} = \sqrt{\gamma^q_{\alpha\alpha'}}
  (1+ r^q_{\alpha\alpha'}) =
  {\beta_\alpha\over \beta_{\alpha'}} t^q_{\alpha'\alpha}\, ,
\label{e96}
\end{equation}
(for details, see \cite{52}). 

It should be pointed out that -- in contrast to the usually
considered external current -- the noise current $\hat{\underline{\bf j}}$ 
as given in Eq.~(\ref{e17}) jumps at the interface, i.e., at 
$z$ $\!=$ $\!0$, which obviously implies the existence of a surface 
noise charge density, and hence the normal component of 
$\epsilon\hat{\underline{\bf E}}$ is not continuous at $z$ $\!=$ $\!0$.
For more details and a derivation of the fields 
$\hat{\underline{\bf E}}$ and $\hat{\underline{\bf B}}$ by direct solution
of the Maxwell equations, the reader is referred to App.~\ref{secapp2}.


\subsection{Commutation Relations}
\label{sec4b}

As shown in App.~\ref{secapp3}, the functions 
$R_{ij}^{\alpha}({\bf r},{\bf r}',\omega)$
and $T_{ij}^{\alpha\alpha'}({\bf r},{\bf r}',\omega)$ 
[Eqs.~(\ref{e78}) -- (\ref{e93})]
obey the relations
\begin{eqnarray}
\int_{-\infty}^{\infty} d\omega \, {\omega\over c^2}
 R^\alpha_{ij} ({\bf r},{\bf r}',\omega)
&=&
\partial^r_i \partial^{r'}_j \tilde{R}^\alpha({\bf r},{\bf r}'),
\label{e98}\\
\int_{-\infty}^{\infty} d\omega \, {\omega\over c^2}
 T^{\alpha\alpha'}_{ij} ({\bf r},{\bf r}',\omega)
&=&
\partial^r_i \partial^{r'}_j 
  \tilde{T}^{\alpha\alpha'}({\bf r},{\bf r}'),
\label{e99}
\end{eqnarray}
where
\begin{eqnarray}
\lefteqn{
\tilde{R}^\alpha({\bf r},{\bf r}')
=
  \int {d^2{\bf k}_\parallel\over (2\pi)^2} \,
  e^{ i{\bf k}_\parallel\cdot({\bf r}_\parallel-{\bf r}_\parallel') }
}
\nonumber \\ && \hspace{10ex}
 \times\,\int_{-\infty}^{\infty} d\omega \, {\omega\over c^2}
  {i\over 2\beta_\alpha} e^{i\beta_\alpha(|z|+|z'|)}
  {r^p_{\alpha\alpha'} \over q_\alpha^2} \, ,
\label{e100}
\end{eqnarray}
\begin{eqnarray}
\lefteqn{
\tilde{T}^{\alpha\alpha'}({\bf r},{\bf r}') 
= 
  \int {d^2{\bf k}_\parallel\over (2\pi)^2} \,
  e^{ i{\bf k}_\parallel\cdot({\bf r}_\parallel-{\bf r}_\parallel') }
}
\nonumber \\ && \hspace{5ex}
  \times\,\int_{-\infty}^{\infty} d\omega \, {\omega\over c^2}
  {i\over 2\beta_\alpha} e^{i\beta_{\alpha}|z|+i\beta_{\alpha'}|z'|}
  \left( 
  - {t^p_{\alpha\alpha'} \over q_\alpha q_{\alpha'}}
  \right).
\label{e101}
\end{eqnarray}
Further it can be shown that
\begin{eqnarray}
\lefteqn{
\int_{-\infty}^{\infty} d\omega \, {\omega\over c^2}
 G^\alpha_{ij} ({\bf r},{\bf r}',\omega)
}
\nonumber \\ && \hspace{5ex}
= i\pi\delta_{ij}\delta({\bf r}-{\bf r}') 
- \partial^r_i\partial^{r'}_j
\tilde{G}^\alpha({\bf r},{\bf r}'),
\label{e102}
\end{eqnarray}
where we have used the Weyl expansion of 
$g^\alpha(|{\bf r}$ $\!-$ $\!{\bf r}'|,\omega)$ \cite{52} to obtain 
\begin{eqnarray}
\lefteqn{
\tilde{G}^\alpha({\bf r},{\bf r}') 
}
\nonumber \\ && \hspace{2ex}
= \int {d^2{\bf k}_\parallel\over (2\pi)^2} \,
  e^{ i{\bf k}_\parallel\cdot ({\bf r}_\parallel-{\bf r}_\parallel') }
  \int_{-\infty}^{\infty} d\omega \, {\omega\over c^2}
  {ie^{i\beta_\alpha|z-z'|}\over {2\beta_\alpha q_\alpha^2}} \, .
\label{e103}
\end{eqnarray}
When we substitute in Eq.~(\ref{e29}) for $G_{ij}({\bf r,r'},\omega)$ 
the actual Green function (\ref{e74}), take advantage of the
relations (\ref{e98}), (\ref{e99}), and (\ref{e102}), and recall
that $\epsilon_{kmj}\partial^{r'}_m \partial^{r'}_j(...)$ $\!=$ $\!0$,
then we can readily prove that the fundamental QED commutation relation 
\begin{eqnarray}
\Bigl[ \hat{E}_i({\bf r}),\hat{B}_k({\bf r}') \Bigr] =
-{i\hbar\over \epsilon_0} \epsilon_{ikm} \partial^{r}_m 
\delta({\bf r}-{\bf r}')
\label{e107}
\end{eqnarray}
is preserved.

In order to prove that the commutations relations between the potentials 
are correct, we first write Eqs.~(\ref{e45}) and (\ref{e47}) [together
with Eqs.~(\ref{e45a}) and (\ref{e47a})] as 
\begin{eqnarray}
\Bigl[
  \hat{A}_i({\bf r}),\,\dot{\!\hat{A}}_j({\bf r}') \Bigr] 
& = &
I_{ij}^{(1)}({\bf r,r'}) ,
\label{e45b} \\
\Bigl[
\hat\varphi({\bf r}),\hat{A}_j({\bf r}')\Bigr] 
& = &
\partial_i^r I_{ij}^{(2)}({\bf r,r'}) ,
\label{e47b}
\end{eqnarray}
where
\begin{eqnarray}
\lefteqn{
I_{ij}^{(k)}({\bf r},{\bf r}') = 
{\hbar\over \pi\epsilon_0}
\int d^3{\bf s} \int d^3{\bf s}' \,
\chi_{im}^{(k)}({\bf r}-{\bf s})
}
\nonumber \\ && \hspace{5ex} \times
  \int_{-\infty}^\infty d\omega {\omega\over c^2} 
   G_{mn}({\bf s}, {\bf s}',\omega)
\delta^\perp_{nj}({\bf s}'-{\bf r}'),
\label{e108}
\end{eqnarray}
with 
\begin{eqnarray}
\label{e108a}
\chi_{im}^{(1)}({\bf r}-{\bf s}) = \delta^\perp_{im}({\bf r}-{\bf s})
\end{eqnarray}
and
\begin{eqnarray}
\label{e108b}
\chi_{im}^{(2)}({\bf r}-{\bf s})=
\frac{\delta_{im}}{4\pi |{\bf r}-{\bf s}|} \, . 
\end{eqnarray}
We use the Green function (\ref{e74}) and take again advantage of
Eqs.~(\ref{e98}), (\ref{e99}), and (\ref{e102}) to derive 
\begin{eqnarray}
\lefteqn{
I_{ij}^{(k)}({\bf r},{\bf r}') = 
{\hbar\over \pi\epsilon_0}
  \sum_{\alpha=1}^{2} \int_{V_\alpha} d^3{\bf s} \,
  \chi_{im}^{(k)}({\bf r}-{\bf s}) 
}
\nonumber \\  && \hspace{2ex} \times  \,
\Biggl\{  
  \int_{V_{\alpha}} d^3{\bf s}' \,
\biggl[ 
    i\pi \delta_{mn}\delta({\bf s}-{\bf s}')
\nonumber \\ && \hspace{2ex}    
       + \partial^s_m \partial^{s'}_n
       \left(-\tilde{G}^\alpha({\bf s},{\bf s}') \! + \!
         \tilde{R}^\alpha({\bf s},{\bf s}')\right)
 \biggr]
\delta^\perp_{nj}({\bf s}'\!-\!{\bf r}')
\nonumber\\  && \hspace{2ex}      
 + \! \int_{V_{\alpha'}} \!\!\! d^3{\bf s}'
    \left[ \partial^s_m \partial^{s'}_n
    \tilde{T}^{\alpha\alpha'}({\bf s},{\bf s}') \right]
    \delta^\perp_{nj}({\bf s}'\!-\!{\bf r}')
\Biggr\}
\nonumber\\&&\hspace{2ex}
= {i\hbar\over \epsilon_0} \sum_{\alpha=1}^{2} 
\!\int_{V_\alpha} \! d^3{\bf s} \,
      \chi_{im}^{(k)}({\bf r}-{\bf s}) \delta^\perp_{mj}({\bf s}-{\bf r}')
    + \tilde{I}_{ij}^{(k)}({\bf r},{\bf r}')
\nonumber\\&&\hspace{2ex}    
= {i\hbar\over \epsilon_0} \int d^3{\bf s} \,
      \chi_{im}^{(k)}({\bf r}-{\bf s}) \delta^\perp_{mj}({\bf s}-{\bf r}')
\label{e109}
\end{eqnarray}
[$\alpha'$ $\!=$ $\!2(1)$ for $\alpha$ $\!=$ $\!1(2)$].
The last line follows from the line before last, because of
\begin{eqnarray}
\lefteqn{    
\tilde{I}_{ij}^{(k)}({\bf r},{\bf r}') =  {\hbar\over \pi\epsilon_0}
  \sum_\alpha 
  \int_{V_\alpha} \!\!d^3{\bf s} \, \chi_{im}^{(k)}({\bf r}\!-\!{\bf s})
}  
\nonumber\\  && \hspace{0ex} \times   
\Biggl\{
  \int_{V_{\alpha}}\!\!d^3{\bf s}'\,
\biggl[ 
    \partial^s_m \partial^{s'}_n
       \left( -\tilde{G}^\alpha({\bf s},{\bf s}') \! + \!
         \tilde{R}^\alpha({\bf s},{\bf s}') \right)         
  \biggr] 
      \delta^\perp_{nj}({\bf s}'\!-\!{\bf r}')
      \nonumber\\  && \hspace{0ex} 
+  \int_{V_{\alpha'}} d^3{\bf s}'
    \left[ \partial^s_m \partial^{s'}_n
    \tilde{T}^{\alpha\alpha'}({\bf s},{\bf s}') \right] 
    \delta^\perp_{nj}({\bf s}'-{\bf r}') 
\Biggr\} = 0 
\label{e110}
\end{eqnarray}
(App.~\ref{secapp4}). Recalling the definitions (\ref{e108a})
and (\ref{e108b}) of $\chi_{im}^{(1)}$ and $\chi_{im}^{(2)}$, 
respectively, from Eq.~(\ref{e109}) it is easily seen that
\begin{eqnarray}
I_{ij}^{(1)}({\bf r},{\bf r}') 
& = & {i\hbar\over \epsilon_0} \int d^3{\bf s} \,
\delta^\perp_{im}({\bf r}-{\bf s}) \delta^\perp_{mj}({\bf s}-{\bf r}') 
\nonumber \\
& = & {i\hbar\over \epsilon_0}\,\delta^\perp_{ij}({\bf r}-{\bf r}')
\label{e111}
\end{eqnarray}
and
\begin{eqnarray}
\partial^{r}_{i}I_{ij}^{(2)}({\bf r},{\bf r}') 
= {i\hbar\over \epsilon_0} \int d^3{\bf s} \,
\partial^r_m \frac{\delta^\perp_{mj}({\bf s}-{\bf r}')}
{4\pi |{\bf r}-{\bf s}|} \, = 0 ,
\label{e112}
\end{eqnarray}
i.e., 
\begin{eqnarray}
\label{e111a}
\Bigl[\hat{A}_i({\bf r}),\,\dot{\!\hat{A}}_j({\bf r}') \Bigr] 
= {i\hbar\over \epsilon_0} \, \delta^\perp_{ij}({\bf r}-{\bf r}')
\end{eqnarray}
and
\begin{eqnarray}
\Bigl[\hat\varphi({\bf r}),\hat{A}_j({\bf r}') \Bigr] = 0.
\label{e111b}
\end{eqnarray}
Hence the theory yields the correct equal-time commutation relations
for both the fields and the potentials.


\section{Conclusions}
\label{sec5}

We have developed a quantization scheme for the electromagnetic
field in a spatially varying three-dimensional linear
dielectric which gives rise to both dispersion and absorption.
Based on the classical phenomenological Maxwell equations,  
the dielectric is described in terms of a complex frequency- and
space-dependent permittivity, which satisfies the Kramers-Kronig relations,
and fluctuating current and charge densities are introduced
in order to be consistent with the dissipation--fluctuation theorem. 
The noise current and charge densities can be thought of as arising
from an additional noise polarization in the constitutive equation
between the dielectric displacement vector and the vector of
the electric field strength. The resulting inhomogeneous Maxwell
equations are then transferred to quantum theory, and the
noise polarization is specified such that the fundamental equal-time
commutations relations of QED are preserved. 

   From the inhomogeneous Maxwell equations together with
the boundary conditions at infinity it follows that 
the electromagnetic field operators can be related, through
the dyadic Green function of the classical problem, to a bosonic field
that represents the elementary (energy) excitations of the overall 
system. This integral representation can be regarded as a natural
extension of the familiar mode expansion in free space or in
cavity-like systems with perfectly reflecting walls. 
Vector and scalar potentials are introduced in the
usual way, and their integral representations are
derived,
which can be used   
in order to couple
the electromagnetic field to additional atomic sources embedded
in the medium.
The fundamental equal-time commutation relations are studied,
and it is found that some of them can be calculated without
kowledge of the explicit form of the Green function in order
to prove the consistence of the quantization scheme with QED.
Others require this knowledge, because single
frequency integrals of the Green function remain to be calculated. 

The determination of the Green function is -- similar to 
the determination of the mode functions in a mode-expansion approach --
a purely classical problem. Its solution is very difficult in general,
and only for simple dielectric-body configurations the Green function
has been calculated so far. 
For a homogeneous dielectric the Green function is well known. We have
used it and explicitly shown that the quantization scheme outlined
here yields exactly the fundamental equal-time commutation relations of QED.
We have further shown that earlier results derived in \cite{39,43}
for the transverse part of the electromagnetic field  are contained 
in our theory. 

An example of an inhomogeneous medium for which the Green function
is known is a configuration of two infinitely extended dielectric
bodies with a common planar interface. We have also used this
Green function and explicitly proved that the quantization scheme
is consistent with QED. Again, recent results given in \cite{39,40}
for paraxial light propagation are recognized.
It is worth noting that the contributions to the Green function
which result from the reflections and transmissions at the
interface do not contribute to the equal-time commutation relations.
Since this is expected to be true also for more complicated configurations,
such as multilayer structures, the equal-time commutation relations 
are expected to be preserved also for these configurations.  

The quantization scheme developed in this paper can be regarded
as the basis for studying the interaction between radiation
and atomic systems in the presence of three-dimensional configurations
of dielectric bodies with dispersion and absorption. In this case,
additional atomic sources must be introduced into the
theory, and the coupled equations of motion for the atomic
variables and the electromagnetic-field variables in the 
Green function expansion (similar to the photonic variables in
a mode expansion) must be tried to be solved.  

\section*{Acknowledgment}
We thank T. Gruner and E. Schmidt for helpful discussions. HTD gratefully
acknowledges support from the Deutscher Akademischer Austauschdienst. He 
would also like to thank the Friedrich-Schiller-Universit\"at Jena for 
hospitality.


\appendix

\newbox{\tmpbox}
\savebox{\tmpbox}{\bf\ref{e28a}}
\section{Proof of Eq.~(\usebox{\tmpbox})}
\label{secapp1}

  From Eq.~(\ref{e23}), the equation for $G_{ni}({\bf s}, {\bf r},\omega)$
reads as
\begin{eqnarray}
\lefteqn{
\left[ \partial^s_l \partial^s_n - \delta_{ln} 
\left(\Delta^s+ {\omega^2\over c^2}\epsilon({\bf s},\omega)\right) \right]
G_{ni}({\bf s}, {\bf r},\omega) 
}
\nonumber \\ && \hspace{35ex}
= \delta_{li}\delta({\bf s}-{\bf r}).
\label{A1}
\end{eqnarray}
We multiply Eq.~(\ref{A1}) from the right by 
$G^*_{lj}({\bf s}, {\bf r}',\omega)$ and integrate 
over ${\bf s}$. We derive, on integrating by parts and taking into
account that the Green function vanishes at infinity,
\begin{eqnarray}
\lefteqn{
{\omega^2\over c^2} \int d^3{\bf s} \, \epsilon({\bf s},\omega)
G_{li}({\bf s}, {\bf r},\omega)G^*_{lj}({\bf s}, {\bf r}',\omega)
}
\nonumber \\ && \hspace{2ex} 
= \int d^3{\bf s} 
\left[ \left(\partial^s_l \partial^s_n - \delta_{ln}\Delta^s\right)
G_{ni}({\bf s}, {\bf r},\omega) \right]
G^*_{lj}({\bf s}, {\bf r}',\omega)
\nonumber \\ && \hspace{10ex}
- G^*_{ij}({\bf r}, {\bf r}',\omega).
\nonumber\\ && \hspace{2ex}
= - \int d^3{\bf s}
\left[ \partial^s_n G_{ni}({\bf s}, {\bf r},\omega) \right]
\left[ \partial^s_l G^*_{lj}({\bf s}, {\bf r}',\omega) \right]
\nonumber\\ && \hspace{4ex}
+ \int d^3{\bf s}
\left[ \partial^s_k G_{li}({\bf s}, {\bf r},\omega) \right]
\left[ \partial^s_k G^*_{lj}({\bf s}, {\bf r}',\omega) \right]
\nonumber \\ && \hspace{10ex}
- G^*_{ij}({\bf r}, {\bf r}',\omega).
\label{A2}
\end{eqnarray}
Now we take the complex conjugate of Eq.~(\ref{A2}), make the interchanges 
${\bf r}\leftrightarrow {\bf r}'$ and $i\leftrightarrow j$, 
and subtract the resulting equation from Eq.~(\ref{A2}). In this way
we arrive at 
\begin{eqnarray}
\lefteqn{
{\omega^2\over c^2} \int d^3{\bf s} \,\epsilon_{\rm I}({\bf s},\omega)
G_{li}({\bf s}, {\bf r},\omega)G^*_{lj}({\bf s}, {\bf r}',\omega)
}
\nonumber \\ && \hspace{5ex}
= {1\over 2i}\left[
G_{ji}({\bf r}', {\bf r},\omega) - 
G^*_{ij}({\bf r}, {\bf r}',\omega) \right]. 
\label{A3}
\end{eqnarray}
Recalling the symmetry property \cite{50}
\begin{eqnarray}
G_{ji}({\bf r}', {\bf r},\omega) = G_{ij}({\bf r}, {\bf r}',\omega)
\label{A4}
\end{eqnarray}
and combining Eqs.~(\ref{A3}) and (\ref{A4}) then yields Eq.~(\ref{e28a}).


\section{Direct solution of the Maxwell equations}
\label{secapp2}

In order to directly solve the Maxwell equations (\ref{e9}) -- (\ref{e12})
together with the permittivity in Eq.~(\ref{e73}), i.e., without using
the Green function, we expand $\hat{\underline{\bf E}}({\bf r},\omega)$ as
\begin{eqnarray}
\hat{\underline{\bf E}}({\bf r},\omega) 
= \int {d^2{\bf k}_\parallel\over (2\pi)^2} \, 
\hat{\underline{\bf E}}({\bf k}_\parallel,\omega;z) 
e^{ i{\bf k}_\parallel \cdot {\bf r}_\parallel}
\label{BB1}
\end{eqnarray}
[cf. Eq.~(\ref{e80}) and the comment made there].
Obviously, similar expressions hold for $\hat{\underline{\bf B}}$ 
and $\hat{\underline{\bf j}}$.
Choosing a reference system in which ${\bf k}_\parallel$ is
parallel to the $x$ axis, the Maxwell equations (\ref{e11}) and 
(\ref{e12}) then yield
\begin{eqnarray}
- \partial_z \hat{\underline E}_y  &=& i\omega \hat{\underline B}_x,
\label{BB3}\\
\partial_z \hat{\underline E}_x - ik_\parallel \hat{\underline E}_z
&=& i\omega \hat{\underline B}_y ,
\label{BB4}\\
ik_\parallel \hat{\underline E}_y
&=& i\omega \hat{\underline B}_z,
\label{BB5}\\
- \partial_z \hat{\underline B}_y
&=& 
- {i\omega\over c^2}\epsilon \hat{\underline E}_x 
+ \mu_0 \,\hat{\!\underline j}_x ,
\label{BB6}\\
\partial_z \hat{\underline B}_x - ik_\parallel \hat{\underline B}_z 
&=&
- {i\omega\over c^2}\epsilon \hat{\underline E}_y 
+ \mu_0 \,\hat{\!\underline j}_y ,
\label{BB7}\\
ik_\parallel \hat{\underline B}_y
&=& 
- {i\omega\over c^2}\epsilon \hat{\underline E}_z 
+ \mu_0 \,\hat{\!\underline j}_z ,
\label{BB8}
\end{eqnarray}
where, for notational convenience, we have omitted the arguments
${\bf k}_\parallel$, $\omega$, and $z$ of the fields 
$\hat{\underline{\bf E}}({\bf k}_\parallel,\omega;z)$, 
$\hat{\underline{\bf B}}({\bf k}_\parallel,\omega;z)$, 
and $\hat{\underline{\bf j}}({\bf k}_\parallel, \omega;z)$, and
we have used the same notation for the rotated reference system
as for the original one. According to the Maxwell equations (\ref{e11}) and 
(\ref{e12}), the tangential components of $\hat{\underline{\bf E}}$ and
$\hat{\underline{\bf H}}$ must be continuous at the surface of discontinuity 
\cite{53a}.

   From Eqs.~(\ref{BB3}), (\ref{BB5}), and (\ref{BB7}) 
together with the boundary conditions we find that 
$\hat{\underline E}_y$ obeys the equation
\begin{eqnarray}
\partial_z^2 \hat{\underline E}_y + \left({\omega^2\over c^2 }\epsilon 
-k_\parallel^2 \right)\hat{\underline E}_y 
= -i\omega\mu_0 \,\hat{\!\underline j}_y,
\label{BB9}
\end{eqnarray}
and $\hat{\underline E}_y$ and $\partial_z\hat{\underline E}_y$
are continuous at the plane $z$ $\!=$ $\!0$.
$\hat{\underline B}_x$ and $\hat{\underline B}_z$ can then be obtained from 
$\hat{\underline E}_y$, using Eqs.~(\ref{BB3}) and (\ref{BB5}).
Similarly, from Eqs. (\ref{BB4}), (\ref{BB6}), and (\ref{BB8})
together with the boundary conditions it follows that $\hat{\underline B}_y$ 
satisfies
\begin{eqnarray}
\partial_z^2 \hat{\underline B}_y + \left({\omega^2\over c^2 }\epsilon 
-k_\parallel^2 \right)\hat{\underline B}_y 
= ik_\parallel \mu_0 \,\hat{\!\underline j}_z 
- \mu_0 \partial_z  \,\hat{\!\underline j}_x,
\label{BB11}
\end{eqnarray}
and $\hat{\underline B}_y$ and $\epsilon^{-1} 
(\partial_z\hat{\underline B}_y$ $\! +$ $\! \mu_0 \,\hat{\!\underline j}_x)$
are continuous at the plane $z$ $\!=$ $\!0$. Knowing $\hat{\underline B}_y$, 
we can find $\hat{\underline E}_x$ and $\hat{\underline E}_z$ from 
Eqs.~(\ref{BB6}) and (\ref{BB8}). 

Thus, Eqs.~(\ref{BB3}) -- (\ref{BB8}) together with the boundary
conditions at infinity and the conditions of continuity of 
the tangential components of $\hat{\underline{\bf E}}$ and
$\hat{\underline{\bf H}}$ at the interface 
provide us with unique solutions for $\hat{\underline{\bf E}}$ and
$\hat{\underline{\bf B}}$. Solving Eqs.~(\ref{BB9}) and (\ref{BB11})
under the conditions mentioned (and going back to the original
reference system) yields the fields $\hat{\underline{\bf E}}$
and $\hat{\underline{\bf B}}$ in full agreement with  
Eqs.~(\ref{e22}) and (\ref{e22a}) and the Green
function from Sec.~\ref{sec4a}.
It should be emphasized that, as can be seen from Eq.~(\ref{BB8}), 
the normal component of $\epsilon\hat{\underline{\bf E}}$ is not 
continuous at the plane $z$ $\!=$ $\!0$, because $\hat{\underline B}_y$ 
is continuous and $\,\hat{\!\underline j}_z$ makes a jump at $z$ $\!=$ $\!0$ 
[see Eq.~(\ref{e17}) together with Eq.~(\ref{e73})].

In the derivation of the Green function in \cite{49} it is stated 
that the continuity of the normal component of 
$\epsilon\hat{\underline{\bf E}}$ at the plane $z$ $\!=$ $\!0$ is 
chosen as a boundary condition. However, one can verify
that it is the continuity of $\epsilon G_{zi}$, $i$ $\!=$ $\!x,y,z$, 
that actually matters, and this can be deduced from the continuity of the
tangential components of $\hat{\underline{\bf E}}$ and 
$\hat{\underline{\bf H}}$. From the continuity of $\epsilon G_{zi}$
it does not follow that $\epsilon\hat{\underline E}_z$ is continuous, 
though the reverse is true. For example, when (as in our case)
$\epsilon G_{zz}$ contains a term proportional to 
$\delta(z$ $\!-$ $\!z')$, which is continuous at the plane 
$z$ $\!=$ $\!0$, then $\epsilon\hat{\underline E}_z$ jumps there,
because it contains a term proportional to  
$\,\hat{\!\underline j}_z$, which makes a jump at $z$ $\!=$ $\!0$ 
[see Eq.~(\ref{e17}) together with Eq.~(\ref{e73})]. 


\savebox{\tmpbox}{\bf Eqs.~(\ref{e98}) and (\ref{e99})}
\section{Proof of \usebox{\tmpbox}}
\label{secapp3}

In order to prove Eqs.~(\ref{e98}) and (\ref{e99}) for
$R^\alpha_{ij}({\bf r,r'},\omega)$ and 
$T^{\alpha\alpha'}_{ij}({\bf r,r'},\omega)$, it is sufficient
to prove them for the Fourier components 
$R^\alpha_{ij}({\bf k}_\parallel,\omega;z,z')$ and 
$T^{\alpha\alpha'}_{ij}({\bf k}_\parallel,\omega;z,z')$. From
Eqs.~(\ref{e81}) -- (\ref{e86}) and Eqs.~(\ref{e87}) -- (\ref{e93}),
respectively, the functions $R^\alpha_{ij}({\bf k}_\parallel,\omega;z,z')$
and $T^{\alpha\alpha'}_{ij}({\bf k}_\parallel,\omega;z,z')$ are seen
to consist, in general, of two parts. One part is associated with the 
$p$-polarized and the other part is associated with the $s$-polarized 
electromagnetic field. Using contour integral techniques (cf. \cite{39,40}), 
it can be 
proved that the latter does not contribute to the integrals on the 
left-hand sides in Eqs.~(\ref{e98}) and (\ref{e99}), 
so that we are left with the contributions from the $p$-polarized 
field only.

Let us consider, e.g., $R^\alpha_{xx}({\bf k}_\parallel,\omega;z,z')$,
Eq.~(\ref{e81}). From
\begin{eqnarray}
\lefteqn{
\int_{-\infty}^\infty d\omega \,{\omega\over c^2} 
{i\over 2\beta_\alpha} e^{i\beta_\alpha(|z|+|z'|)}
   {r^p_{\alpha\alpha'} \over q_\alpha^2} 
  \left(-\beta_\alpha^2 \right) {k_x^2\over k_\parallel^2} 
}
\nonumber \\ && \hspace{2ex}  
= \int_{-\infty}^\infty\!\! d\omega \,{\omega\over c^2} 
{i\over 2\beta_\alpha} e^{i\beta_\alpha(|z|+|z'|)}
   {r^p_{\alpha\alpha'} \over q_\alpha^2} 
  \left( - q_\alpha^2 \!+\! k_\parallel^2 \right) {k_x^2\over k_\parallel^2} 
\nonumber\\ && \hspace{2ex}
= \int_{-\infty}^\infty d\omega \,{\omega\over c^2} 
{i\over 2\beta_\alpha} e^{i\beta_\alpha(|z|+|z'|)}
   {r^p_{\alpha\alpha'} \over q_\alpha^2} k_x^2
\label{B1}
\end{eqnarray}
Eq.~(\ref{e98}) can readily be proved correct for 
$R^\alpha_{xx}({\bf r},{\bf r}',\omega)$. In exactly the same way, 
one can show that Eq.~(\ref{e98}) also holds for the remaining 
functions $R^\alpha_{ij}({\bf r},{\bf r}',\omega)$.
In order to show that Eq.~(\ref{e99}) is valid, it is helpful to 
perform the integration over $\omega$ in Eq.~(\ref{e101}) explicitly.
Using the relation
\begin{eqnarray}
e^{i\beta_{\alpha}|z|}= \int_{-\infty}^\infty {dk\over 2\pi} \, e^{ikz}
{ 2i\beta_{\alpha}\over \beta_{\alpha}^2 - k^2 } \, ,
\label{B2}
\end{eqnarray}
we have 
\begin{eqnarray}
\lefteqn{
\int_{-\infty}^\infty d\omega \, {\omega\over c^2}
{i\over 2\beta_\alpha} e^{i\beta_{\alpha}|z|+i\beta_{\alpha'}|z'|}
   \left( - {t^p_{\alpha\alpha'} \over q_\alpha q_{\alpha'}} \right)
}   
\nonumber\\ && \hspace{2ex}
= \int_{-\infty}^\infty {dk\over 2\pi} \, e^{ikz} 
\int_{-\infty}^\infty {dk'\over 2\pi} \, e^{ik'z'}
\nonumber\\ && \hspace{2ex} \times
\int_{-\infty}^\infty d\omega \, {\omega\over c^2}
  {i\over 2\beta_\alpha}
  { 2i\beta_{\alpha}\over \beta_\alpha^2 - k^2 }
  { 2i\beta_{\alpha'}\over \beta_{\alpha'}^2 - {k'}^2 }
  \left( - {t^p_{\alpha\alpha'} \over q_\alpha q_{\alpha'}} \right).
\label{B3}
\end{eqnarray}
Since the equation $\beta_\alpha^2-k^2$ $\!=$ $\!0$ has no solutions 
in the upper complex frequency half-plane \cite{37,40}, 
the integrand in the $\omega$ integral in Eq.~(\ref{B3}) has no 
poles there. Treating the single pole at $\omega$ $\!=$ $\!0$ 
as principal value, we obtain
\begin{eqnarray}
\lefteqn{
\int_{-\infty}^\infty d\omega \, {\omega\over c^2}
{i\over 2\beta_\alpha} e^{i\beta_{\alpha}|z|+i\beta_{\alpha'}|z'|}
   \left( - {t^p_{\alpha\alpha'} \over q_\alpha q_{\alpha'}} \right)
}
\nonumber \\ && \hspace{2ex}   
= \int_{-\infty}^\infty {dk\over 2\pi} \,
  {e^{ikz} \over k_\parallel^2 + k^2}
\int_{-\infty}^\infty {dk'\over 2\pi} 
  {e^{ik'z'} \over k_\parallel^2 + {k'}^2}
\nonumber \\ && \hspace{10ex} \times \,  
  2i\pi k_\parallel 
  \left( - 
     { t^p_{\alpha\alpha'}(0) \over \sqrt{ \epsilon_\alpha(0) \epsilon_{\alpha'}(0)
     }}
  \right).  
\label{B4}
\end{eqnarray}
Let us now turn to the left-hand side in Eq.~(\ref{e99}) and
consider, e.g., $T^{\alpha\alpha'}_{xx}({\bf k}_\parallel,\omega;z,z')$,
Eq.~(\ref{e87}). We derive 
\begin{eqnarray}
\lefteqn{
\int_{-\infty}^\infty d\omega \, {\omega\over c^2}
{i\over 2\beta_\alpha} e^{i\beta_{\alpha}|z|+i\beta_{\alpha'}|z'|}
  {t^p_{\alpha\alpha'} \over q_\alpha q_{\alpha'}}
  {\beta_\alpha \beta_{\alpha'}\over k_\parallel^2}
}
\nonumber \\ && \hspace{1ex}
=  \int_{-\infty}^\infty {dk\over 2\pi} \, e^{ikz} 
\int_{-\infty}^\infty {dk'\over 2\pi} \, e^{ik'z'}
\int_{-\infty}^\infty d\omega \, {\omega\over c^2}
\nonumber \\ && \hspace{4ex}
\times  {i\over 2\beta_\alpha} 
{ 2i\beta_{\alpha}\over \beta_\alpha^2 - k^2 }
  { 2i\beta_{\alpha'}\over \beta_{\alpha'}^2 - {k'}^2 }
  {t^p_{\alpha\alpha'} \over q_\alpha q_{\alpha'}} 
 {\beta_\alpha \beta_{\alpha'}\over k_\parallel^2} 
\nonumber \\ && \hspace{1ex}
= 
\int_{-\infty}^\infty {dk\over 2\pi} \, 
  {e^{ikz} \over k_\parallel^2 + k^2}
\int_{-\infty}^\infty {dk'\over 2\pi} \,
  {e^{ik'z'} \over k_\parallel^2 + {k'}^2}
\nonumber\\&& \hspace{8ex}
\times 2i\pi k_\parallel
\left( -
  {t^p_{\alpha\alpha'}(0) \over \sqrt{ \epsilon_\alpha(0) \epsilon_{\alpha'}(0) }}
\right)\,,
\label{B5}
\end{eqnarray}
which exactly agrees with Eq.~(\ref{B4}), i.e., Eq.~(\ref{e99}) is
proved correct for $T^{\alpha\alpha'}_{xx}({\bf r},{\bf r}',\omega)$.
Eq.~(\ref{B4}) for the other functions 
$T^{\alpha\alpha'}_{ij}({\bf r},{\bf r}',\omega)$
can be given in the same way. 


\savebox{\tmpbox}{\bf\ref{e110}}
\section{Proof of Eq.~(\usebox{\tmpbox})}
\label{secapp4}

We perform in Eq.~(\ref{e110}) a partial integration over $s'$ and
obtain, on recalling that 
\mbox{$\partial^{s'}_n$ $\! \delta^\perp_{nj}({\bf s}'$ $\!-$ $\!{\bf r}')$}
$\!=$ $\!0$,
\begin{eqnarray}
\lefteqn{
\tilde{I}^{(k)}_{ij}({\bf r},{\bf r}') 
} 
\nonumber\\ && \hspace{2ex} 
=  -{\hbar\over \pi\epsilon_0}
 \sum_\alpha \int_{V_\alpha} d^3{\bf s} \, \chi^{(k)}_{im}({\bf r}-{\bf s})
 \int ds'_x ds'_y 
\nonumber\\ && \hspace{5ex} \times \,
 \partial^s_m
 \left[
       -\tilde{G}^\alpha({\bf s},{\bf s}') 
       + \tilde{R}^\alpha({\bf s},{\bf s}')
 \right.      
\nonumber\\ && \hspace{10ex}        
 \left.
       - \tilde{T}^{\alpha\alpha'}({\bf s},{\bf s}')
 \right] 
 \delta^\perp_{nj}({\bf s}'-{\bf r}') \biggr|_{s'_z=0}.
\label{C1}
\end{eqnarray}
Using Eqs.~(\ref{e100}), (\ref{e101}), and (\ref{e103}), we derive
\begin{eqnarray}
\left[-\tilde{G}^\alpha({\bf s},{\bf s}') 
+ \tilde{R}^\alpha({\bf s},{\bf s}')
- \tilde{T}^{\alpha\alpha'}({\bf s},{\bf s}') \right]\biggr|_{s'_z=0} 
\nonumber\\
{\lefteqn =\quad
\int {d^2{\bf k}_\parallel\over (2\pi)^2}
  e^{ i{\bf k}_\parallel\cdot({\bf s}_\parallel-{\bf s}_\parallel') }
  \int_{-\infty}^{\infty} d\omega {\omega\over c^2}
  {i\over 2\beta_\alpha} 
}
\nonumber\\   
 \times e^{i\beta_\alpha|s_z|}
  \left( - {1\over q_\alpha^2}
         + {r^p_{\alpha\alpha'} \over q_\alpha^2}
         + {t^p_{\alpha\alpha'} \over q_\alpha q_{\alpha'}}
  \right)
\nonumber\\
{\lefteqn =\quad
\int {d^2{\bf k}_\parallel\over (2\pi)^2}
  e^{ i{\bf k}_\parallel\cdot({\bf s}_\parallel-{\bf s}'_\parallel) }
  \int_{-\infty}^{\infty} d\omega {\omega\over c^2}
  {i\over 2\beta_\alpha} 
}
\nonumber\\  
\times  e^{i\beta_\alpha|s_z|}
  \left( - {1\over q_\alpha^2}
         + {r^p_{\alpha\alpha'} \over q_\alpha^2}
         + {\beta_\alpha\over\beta_{\alpha'}}
           \sqrt{\epsilon_{\alpha'} \over \epsilon_\alpha}
           {1- r^p_{\alpha\alpha'} \over q_\alpha q_{\alpha'}}
  \right)
\nonumber\\
{\lefteqn =\quad
\int {d^2{\bf k}_\parallel\over (2\pi)^2}
  e^{ i{\bf k}_\parallel\cdot({\bf s}_\parallel-{\bf s}'_\parallel) }
  \int_{-\infty}^{\infty} d\omega {\omega\over c^2}
  {i\over 2\beta_\alpha} 
}
\nonumber\\  
\times e^{i\beta_\alpha|s_z|}
  {1- r^p_{\alpha\alpha'} \over q_\alpha^2}
  \left( - 1
         + {\beta_\alpha\over\beta_{\alpha'}}
  \right)
= 0. 
\label{C2}
\end{eqnarray}
To obtain the second equation from the first one, we have used
Eq.~(\ref{e96}) for $t^p_{\alpha\alpha'}$. The third equation gives zero
because of the integration over $\omega$. Combining Eqs.~(\ref{C2}) 
and Eq.~(\ref{C1}) yields Eq.~(\ref{e110}).

\vspace {0.6cm}

\noindent
*Permanent adress: Institute of Physics, National Center
for Sciences and Technology, 1 Mac Dinh Chi St., Dist. 1, Ho Chi Minh city,
Vietnam.

\end{document}